\documentclass[prl,twocolumn]{revtex4-1}
\usepackage{graphicx}
 \usepackage{amsmath}

\begin{document}

\title{Non-reciprocal neutral ferroelectric domain walls in BiFeO$_3$}

\author{M. A. P. Gon\c{c}alves,  M. Graf, M. Pa\'sciak, and J. Hlinka}
\thanks{Corresponding author. Email: hlinka@fzu.cz}
\affiliation{FZU - Institute of Physics of the Czech Academy of Sciences\\%
Na Slovance 2, 182 21 Prague 8, Czech Republic}

\date{\today}

\begin{abstract}
This paper analyzes a peculiar phenomenon of non-reciprocity of domain walls and illustrates its implications using ab-initio-based atomistic computational experiments with ($\bar{1}\bar{1}2$)-oriented planar 180$^{\circ}$-domain walls within the  canonical multiferroic ferroelectric crystal of BiFeO$_3$.
Results show that walls on the opposite sides of a given domain lamellae within a twinned domain structure can have considerably different properties, such as different polarization and oxygen octahedra tilt profiles, different thicknesses and different wall energy densities. 
The spontaneous formation of 180$^{\circ}$ zigzag walls and triangular domains of inverted polarization suggests that one of the non-reciprocal ($\bar{1}\bar{1}2$)-oriented domain walls is actually the lowest energy 180$^{\circ}$ wall of the pure insulating BiFeO$_3$.
\end{abstract}




\maketitle 



 Domain structure is a natural manifestation of symmetry breaking phase transition in a crystal.
 A frequent domain pattern is the lamellar twinning, formed by alternation of two distinct domains states, $A$ and $B$, and separated by parallel planar domain walls.
 Often, the two subsequent walls $A|B$ and $B|A$ are equivalent by symmetry reasons.
 %
 However, it is also possible that the symmetry relationship between A$|$B and B$|$A is absent.
 Then, $A|B$ and $B|A$ are physically non-equivalent walls. 
 Such non-reciprocal walls were earlier also referred as irreversible ones~\cite{Janovec81,Janovec}.

  
 Any nominally {\it  charged} ferroelectric domain wall is a non-reciprocal wall.
In a lamellar structure, the subsequent domains usually have an opposite normal component of the spontaneous polarization and subsequent domain walls then differ by the sign of the interfacial charge.
  They can be thus denoted as head-to-head (HH) or tail-to-tail (TT) walls, respectively, and the absence of symmetry relation is obvious.
  %
  %

  In this paper, we are addressing the question whether and when one can encounter also manifestations of the non-reciprocity of nominally {\it  neutral} ferroelectric  walls.
  In particular, we explore the case of the so-called \cite{Marton2010}
  R180$^{\circ}$($\bar{1}\bar{1}2$) walls 
  and argue that these in BiFeO$_3$ experimentally reported \cite{experiment}
  but so far little investigated walls are indeed showing remarkable non-reciprocal effects.

\begin{figure}[ht]
\centering
\includegraphics[width=.7\columnwidth]{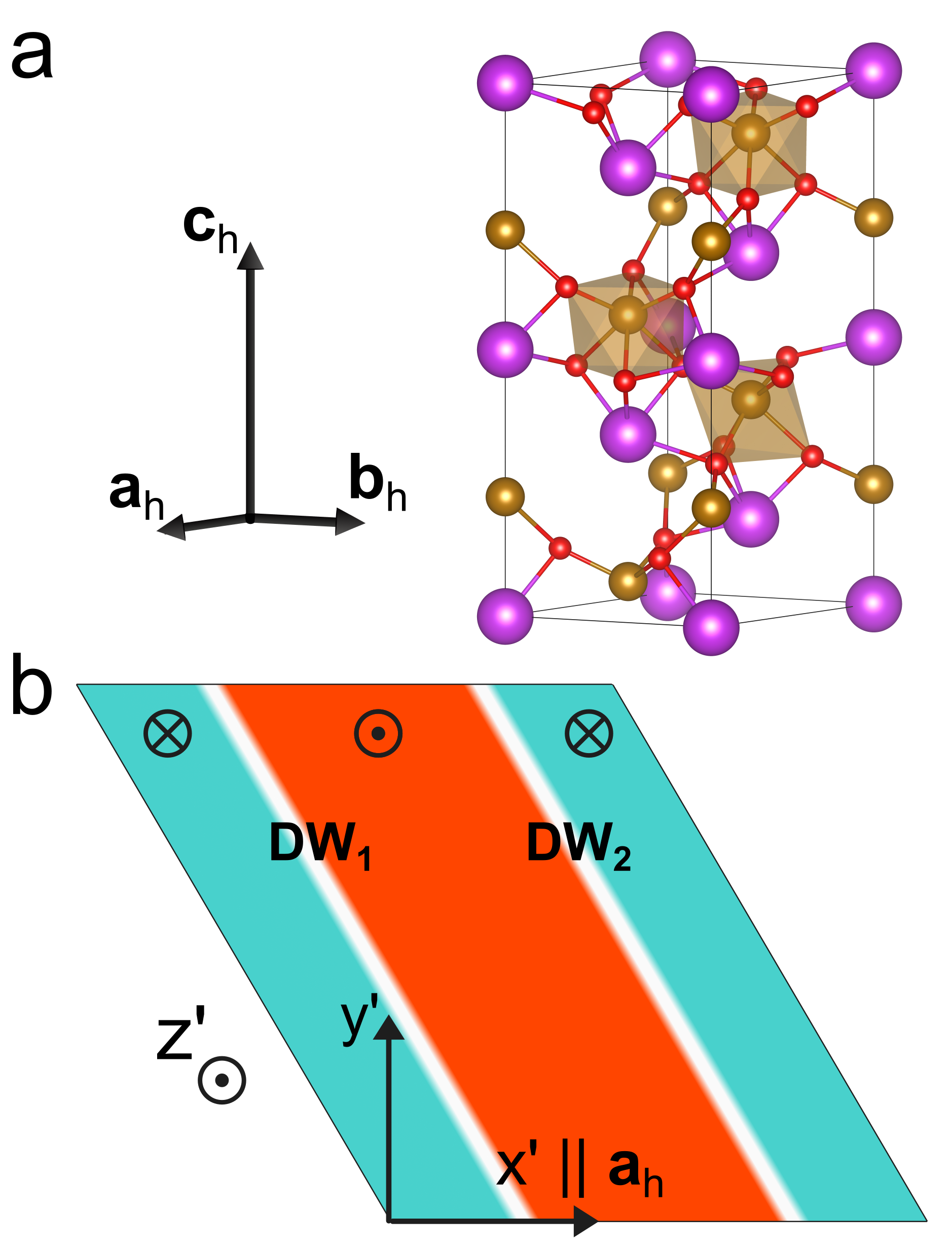}
\caption{ BiFeO$_3$ structure and the orientation of the investigated R180($\bar{1}2\bar{1}$) domain walls. (a) hexagonal cell and the corresponding lattice vectors. Purple spheres stand for Bi, red spheres for O, dark yellow for Fe. (b) Schema of the simulation box with two  domain walls, parallel to the plane defined by $\bf a_{\rm h}$ and $\bf b_{\rm h}$ lattice vectors.  Central (orange) domain stripe has polarization in a positive sense of $z'$ axis,  cyan domains have polarization in the opposite sense. Adopted Cartesian coordinate system $x'$, $y'$ and $z'$ is also indicated.
}
\label{fig1}
\end{figure}

BiFeO$_3$ is a well-studied multiferroic material with coexisting  ferroelectric, ferroelastic, antiferrodistortive and antiferromagnetic orders and promising applications in spintronics, nonvolatile memory devices, catalysis and sensors~\cite{wang_2020, chaudron_2024,YangAbove, Amdouni,Huyan2019}.
At room temperature, it has rhombohedral $R3c$ symmetry with a spontaneous polarization of about 1\,C/m$^2$ and with oxygen octahedra alternatively tilted by about $14^\circ~$ around the polarization axis~\cite{lebeugle_2007,Catalan09}.
The conventional hexagonal unit cell is depicted in Fig.\,1a.
%
When neglecting the spontaneous strain, its lattice vectors can be expressed in the basis of the parent high-temperature $Pm\bar{3}m$ phase as ${\bf a_{\rm h}}=(0,1,\bar{1})$, ${\bf b_{\rm h}}=(\bar{1}, 0, 1)$ and ${\bf c_{\rm h}}=(2,2,2)$.
This relation allows us to describe crystallographic directions with Miller indices $h, k, l$ of the parent cubic crystal structure.
It also implies the existence of the eight equivalent ferroelectric domain states with polarization along $\langle 111 \rangle$ cubic diagonals.
Corresponding ferroelectric domain walls are primarily distinguished as 180$^\circ$, 109$^\circ$ and 71$^\circ$ ones according to the angles between the polarization in the adjacent domains.
Structure and properties of domain walls in BiFeO$_3$ have been subject of multiple experimental~\cite{Farokhipoor,Sando14,Bhatnagar,Geng2020,Chauleau2020,Fusil2022,Condurache} and theoretical studies~\cite{Dieguez2013, Ren2013, ourPRL, Chauleau2020, Lubk, Wang2013, Xue2014, Zhang2022, Mangeri, Eliseev}.
The benchmark systematic first-principles study of Ref.\,\onlinecite{Dieguez2013} explored almost all types of mechanically and electrically compatible (neutral) domain walls.
However, non-reciprocal walls were not addressed there.

To demonstrate and explain several remarkable non-reciprocal domain wall effects 
in 180$^\circ~$ domain walls of BiFeO$_3$
we consider two antiparallel domain states with polarization along  $[111]$ and  $[\bar{1}\bar{1}\bar{1}]$. 
The mechanically and electrically compatible 180$^\circ~$  domain walls connecting these two states should be parallel to the $[111]$ direction.
%
There are two high-symmetry options complying with this requirement.
One option is realized by $(1\bar{1}0)$-type planes, {\it i.e.} walls parallel to the mirror symmetry plane of the parent $Pm\bar{3}m$ reference phase.
For example, it would be $y'z'$ plane within the Cartesian coordinate system $x'$, $y'$ and $z'$ with $x'\parallel {\bf a_{\rm h}}$ and $z'\parallel {\bf c_{\rm h}}$ shown in Fig.\,1.
The other option represents the ($\bar{1}\bar{1}2$) plane, passing through intersecting 3-fold and 2-fold rotational axes of the parent $Pm\bar{3}m$ structure.
Most of the studies have so far considered only the former type. 
However, as we shall show, the latter ($\bar{1}\bar{1}2$)-type walls, such as DW$_1$ and DW$_2$ depicted in Fig.\,1b, are even more important for BiFeO$_3$.


Our predictions are based on an interatomic potential with parameters fitted to ab-initio calculations~\cite{Graf2014,Graf2014b}. 
This interatomic potential describes each atom with a core and a shell with a mutual anharmonic core-shell interaction mimicking the atomic polarizability~\cite{Tinte2004}.
It has been successfully applied to BiFeO$_3$ in few other studies~\cite{Graf2015,Graf2016,ourPRL}.
%
The domain wall structures were relaxed using classical molecular dynamics simulations under constant stress and temperature conditions, using a supercells with typically
20$\times$20$\times$2 conventional rhombohedral BiFeO$_3$ cells of Fig.\,1a (\textit{i.e.} 24000 atoms) subjected to periodic boundary conditions.
 A temperature of 1~K was chosen to facilitate effective relaxation while introducing a minimal amount of thermal noise. 
 The simulation began with a 10~ps thermalization, followed by an additional 10~ps during which the average configuration was computed.
The molecular dynamics calculations were performed using the DL\_POLY software~\cite{Todorov2006}.
Local dipole moments were evaluated for each 5-atom perovskite cell centered in the Bi atoms, based on the positions and charges of cores and shells.  
For a given cell, one Bi atom with its surrounding twelve O atoms (with a weight of 0.25) and eight coordinating Fe atoms (0.125 weight) were considered.
The desired ($\bar{1} 2 \bar{1}$) domain wall orientation was selected by imposing the suitable step-like distribution of the initial Bi ion off-centering displacements in our simulation box. 
The periodic boundary conditions requires to introduce  two such domain walls, as indicated in  Fig.\,1b. 


After the relaxation, most of the unit cells have their structure similar to that of the homogeneous ground state. 
For the detailed inspection of the region near the domain wall,  polarization of each unit cell was projected on the rotated coordinate system, where 
$s \parallel  [\bar{1}2\bar{1}]$ is perpendicular to the wall, $z' \parallel  [111]$ is along the polarization,  and $t \parallel {\bf b_{\rm h}}$ is perpendicular to the both $s $ and $ z'$.
Resulting polarization profiles in the vicinity of domain wall DW$_1$ and DW$_2$ are shown in Fig.\,2a,b.
Analogous profiles of the staggered oxygen octahedra tilt rotations $R_i$ are shown in Fig.\,2c,d.
As expected,
$P_{\rm s}$ and $P_{\rm t}$ components are zero or close to zero, while the $P_{\rm z'}$ component is switching between the nominal spontaneous polarization values. 
The absence of switching of the $R_{\rm i}$ components confirms that neither DW$_1$ nor DW$_2$  coincide with an anti-phase boundary of the underlying staggered tilt pattern and thus both walls are purely ferroelectric domain walls.

While the profiles of $P_{\rm s}$,  $P_{\rm z'}$, $R_{\rm s}$,  and   $R_{\rm z'}$ 
components are same at any line passing through the given wall,
 the values of $P_{\rm t}$ and $R_{\rm t}$ alternate between two mutually opposite values when comparing lines related by the fractional translation vector ${\bf c_{\rm h}}/2$.
 It means that the Bloch polarization component averaged over the two
lines is zero, but at the same time, there is a staggered (anti-ferroelectric) Bloch component within the wall. 
In case of the tilt profile, the double signs of $R_{\rm t}$ implies that the Bloch component of the staggered oxygen octahedra rotation averaged over the two
 octahedra related by ${\bf c_{\rm h}}/2$ is zero, but both octahedra  are slightly rotated in the same sense around the $t$-axis.

There are also important differences between DW$_1$ and DW$_2$.
First, the DW$_1$ wall thickness defined by slope at the center of the wall as in Refs.\,\onlinecite{Hlinka2006, Marton2010} is about 0.7\,nm, while the DW$_2$ thickness  is only about 0.35\,nm.
Second, the 
magnitude of the tilt modulations is much stronger in DW$_1$.
Third, the signs of the $R_{\rm s}$ are opposite in DW$_1$ and DW$_2$. 
A more detailed analysis of these differences  within a  continuous Ginzburg-Landau theory is a subject of a follow-up paper.

\begin{figure}[ht]
\centering
\includegraphics[width=\columnwidth]{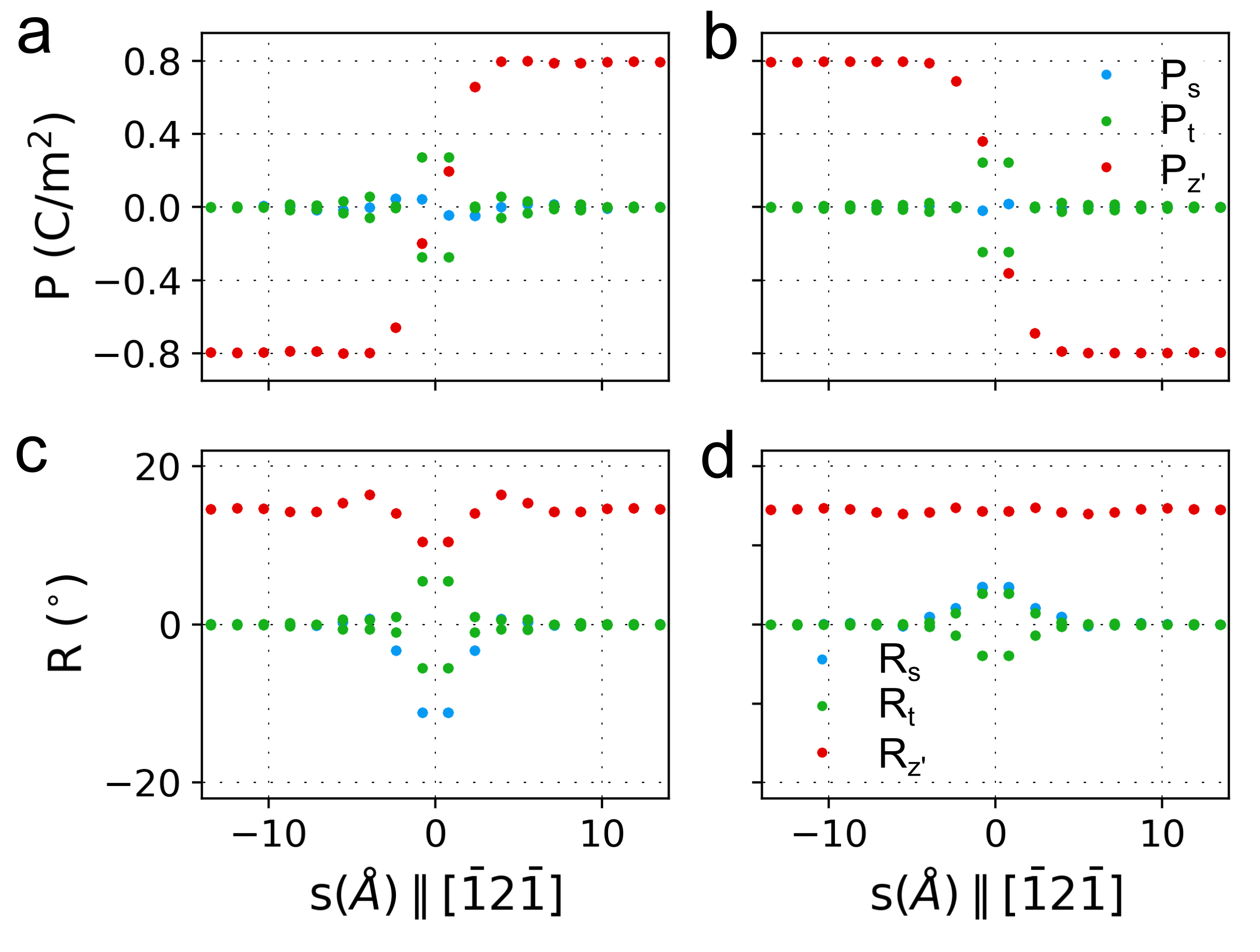}
\caption{Profiles of the components polarization $P_i$ (a,b) and oxygen octahedra tilts $R_i$ (c,d) across the DW$_1$ (a,c) and DW$_2$ (b,d) domain walls defined in Figure 1b. The rotated Cartesian coordinate system $s, t, z'$ is used.
}
\label{fig:Profily}
\end{figure}

To verify that the  profiles of Fig.\,2 are the lowest energy domain wall configurations,
we have slightly varied the initial positions of the walls, size of the simulation box  and we also performed annealing at higher temperatures.
In overall, the domain characteristics described above proved to be robust and representative ones. 

Nevertheless, an interesting phenomenon was encountered in the course of gradual temperature increase in steps of 20\,K up to 300\,K~\cite{extradetails}. 
While the time-averaged polarization map at $z'=0$ was practically the same till  $T = 120$\,K (see Fig.\,3a),
as soon as the temperature reached $T = 140$\,K, the  DW$_2$ started to display significant fluctuations. 
The original planar wall changed to an irregularly corrugated shape and the $P_{\rm t}$ components started to be disordered. 
The slow fluctuations are apparent even in the 10~ps averaged polarization maps (see Fig.\,3b).
Finally, at $T = 240$\,K, the shape of DW$_2$ abruptly transformed  into a zigzag wall with 60$^\circ~$ folds.
This form of DW$_2$ then persists till $T = 300$\,K (see Fig.\,3c) and it remains there even when the system is cooled back down to $T = 1$\,K.


\begin{figure}[ht]
\centering
\includegraphics[width=.8\columnwidth]{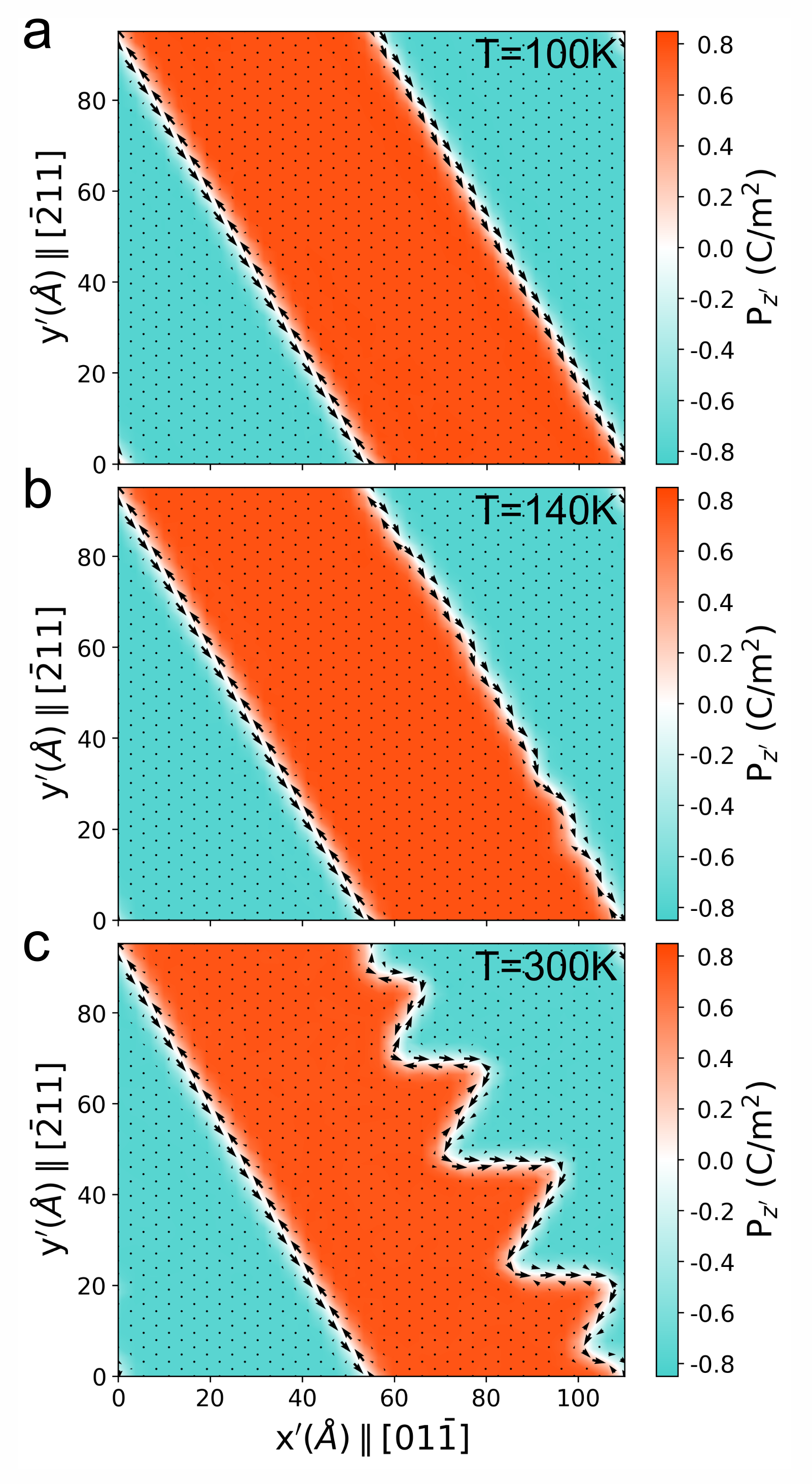}
\caption{Calculated polarization maps showing evolution of the 180$^\circ$ domain walls while increasing the  temperature to (a) $100$\,K, (b)$140$\,K and (c) $300$\,K. Arrows show the non-negligible in-plane polarization components, the color shows the out-of-plane polarization component.
}
\label{fig:Zizag}
\end{figure}

Finally, we also tried to stabilize small diameter columnar nanodomain of inverted polarization.
%
%
The only encountered stable nanodomain had a shape of a  right  prism with a equilateral triangular cross section as shown in Fig.\,4. 
We could see that one of the sides of this triangular prism is always parallel the walls of Fig.\,1b,  and that the wall profiles correspond to that of DW$_1$ in Fig.\,2a,c. 
%
We have never observed a spontaneous formation of
 DW$_2$ walls or $(1\bar{1}0)$-type walls, suggesting that DW$_1$ is actually the lowest energy variant of the 180$^\circ$  domain wall.

Summarizing, we have shown that the DW$_2$ differs from DW$_1$ by the thickness, by the amplitudes of the in-wall order parameters and also by the stability. 
The ensemble of all these findings implies that DW$_1$ and DW$_2$ are not related by any symmetry operation and so these walls are clearly non-reciprocal ones.

\begin{figure}[ht]
\centering
\includegraphics[width=.9\columnwidth]{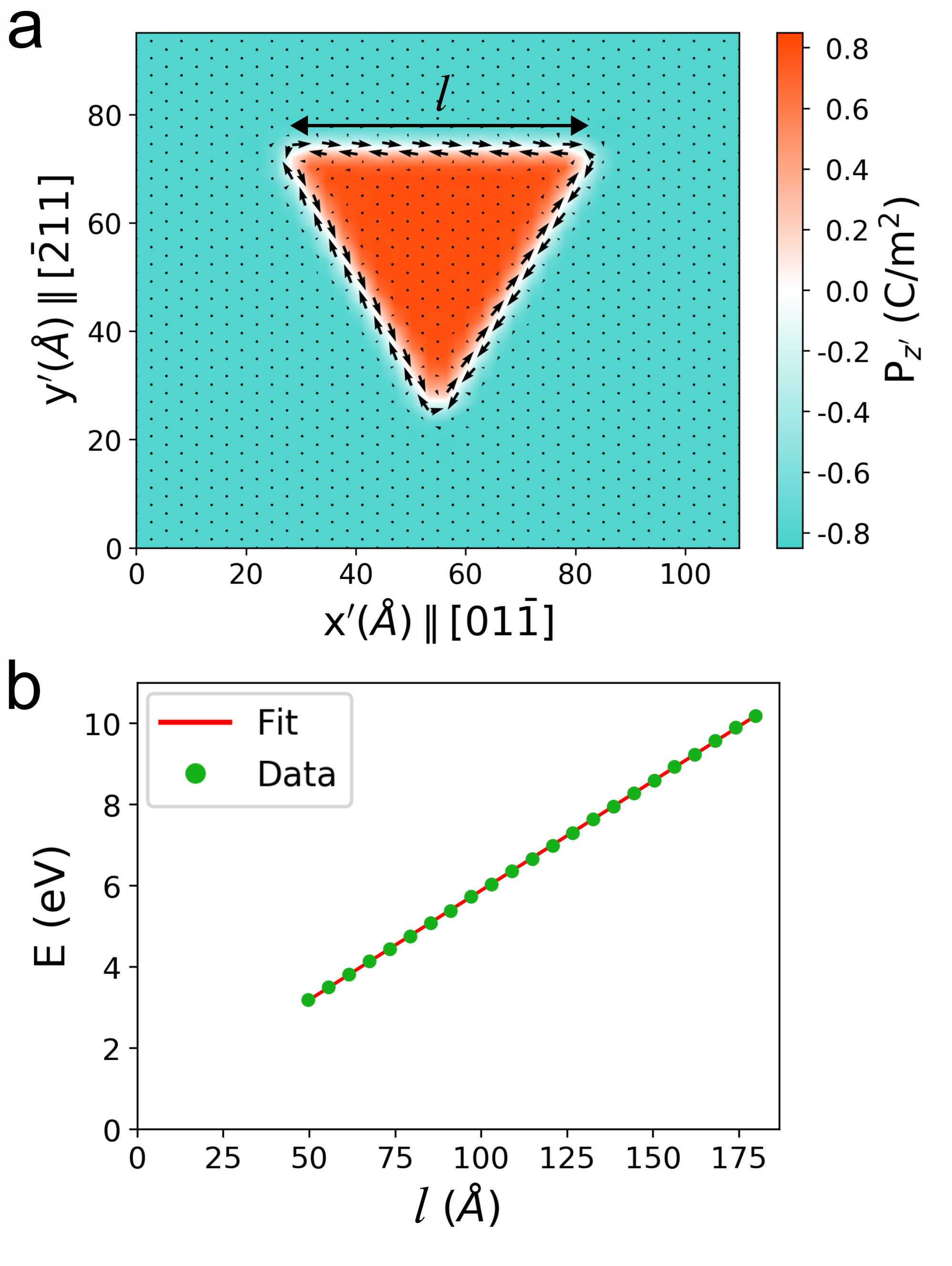}
\caption{ Stable nanodomains of antiparallel polarization.
(a) Polarization map at $z'=0$ cross section through a stable prismatic nanodomain. (b) The excess energy due to the domain walls as a function of wall-fold distance $l$  and the linear fit to the data.
}
\label{fig:Triangle}
\end{figure}


To facilitate the discussion, let us introduce a more precise domain wall symbol
$ [A | {\bf{n}} \rangle B]$.
Here $A$ and $B$ stand for domain states, while the oriented domain wall normal $\bf{n} $ points in the real space towards the domain that is ascribed to the  state of the right-hand side of the symbol -- as it is mnemonically indicated by the embedded ``ket" symbol. 
With the already introduced cubic Miller indices for spontaneous polarization directions and oriented domain wall normals, our DW$_1$ of Fig.\,1b can be
described as  
$[111 | 1\bar{2}1 \rangle \bar{1}\bar{1}\bar{1}]=[\bar{1}\bar{1}\bar{1} | \bar{1}2\bar{1} \rangle 111 ]$, while DW$_2$ can be described as  
$[111 |        \bar{1}2\bar{1}     \rangle \bar{1}\bar{1}\bar{1}]=[\bar{1}\bar{1}\bar{1} | 1\bar{2}1    \rangle 111 ]$.

Symmetry of planar domain walls was systematically developed in Refs.\,\onlinecite{Janovec81,Janovec}.
The basic assumption is that specific symmetries are described by certain subgroups of the parent phase crystallographic group $G$. 
The  symmetry of the $[A | {hkl} \rangle B]$ wall in the usual sense is described by a group $T_{AB}|{hkl}\rangle $, composed of those parent phase symmetry group operations, that generates image structures  indistinguishable from the original domain wall structure.
It is a crystallographic group with two-dimensional translational symmetry.
The highest possible symmetry of the wall $T_{AB}^0(hkl) $ can be determined considering the orientation of the domain wall plane and its centering in the crystal structure~\cite{Janovec81,Janovec,Schranz}.
%
In the present case, the $2_{\rm t}/m_{\rm t}$ factor group of $T_{AB}^0(hkl)$  contains the identity, inversion $i$, mirror $m_{\rm t}$ perpendicular to $t$ axis, and the 2-fold rotation  $2_{\rm t}$ (parallel to the $t$ axis). The profiles of Fig.\,2 agree with this symmetry within the precision of our numerical calculations.

The non-reciprocity phenomenon is related to the symmetry group of domain wall pair $\bar{J}_{AB}(hkl) $, which collects all parent phase symmetry operations that preserve or exchange the domain states and preserve the domain wall plane $(hkl) $.
It is a supergroup of $T_{AB}^0(hkl) $ because it may also contain operations that interchange the sides of the wall but do not alter the domain state (pure side-reversal operations) or the operations that reverse the domain states but do not interchange sides of the wall (pure state-reversal operations).
The existence of such pure side-reversal or pure state-reversal symmetry operations would imply 
the reciprocity (reversibility) of the wall. 
In the case of the present wall, no such symmetry operations exists and $\bar{J}_{AB}(hkl)$ equals to $T_{AB}^0(hkl)$.

Finally,  the group of the unordered domain pair, $J_{AB}$, involves all operations of $G$ maintaining or interchanging the two domain states \cite{Schranz}.
%
In our case the  $\bar{3}m$ factor group of $J_{AB}$ is a nontrivial supergroup of the factor group of $\bar{J}_{AB}|(hkl)$ 
what implies that there are other walls equivalent to DW$_1$ that join the same two domain states.
It can be seen that these are 
the 
$[111 | \bar{2}11 \rangle \bar{1}\bar{1}\bar{1}]=
[\bar{1}\bar{1}\bar{1} | 2\bar{1}\bar{1} \rangle 111 ]$ 
and the
$[111 | 11\bar{2} \rangle \bar{1}\bar{1}\bar{1}]=
[\bar{1}\bar{1}\bar{1} | \bar{1}\bar{1}2  \rangle 111 ]$ walls
corresponding to the segments of the the zigzag wall in Fig.\,3b.
All these three equivalent walls are realized in the three sides of the triangular prism in Fig.\,4a.

The non-reciprocal walls should also differ in their planar energy density. 
The irreversible decomposition of DW$_2$ wall documented in Fig.\,3 and the systematic preference for DW$_1$-type domain walls in stable prismatic nanodomains (Fig.\,4a) suggests that the planar domain wall energy density $\sigma_2$ of DW$_2$ is significantly higher then $\sigma_1$ of DW$_1$. 
The broadly applied  procedure of the  domain wall energy density determination from computational experiments consists of (i) calculation of energy of planar domains with desired crystallographic orientation in a periodic cell like that of Fig.\,1, (ii) subtracting the corresponding energy of the single domain state, and (iii) dividing the difference by the total surface of the walls within the supercell.
In case of non-reciprocal wall pair, however,
this yields only the average energy density $(\sigma_1+\sigma_2)/2$. 
In order to overcome this topological obstacle, 
we have thus used the prismatic nanodomain of Fig.\,4a.
As argued above,  all three sides walls  are equivalent to DW$_1$ wall.
For elimination of the domain fold energy contributions, we have 
calculated a whole sequence of relaxed nanodomain configurations with the wall-fold distance (the arm of the displayed triangle) $l$, ranging from 5 to 20\,nm. 
The slope of the excess energy associated with the domain walls {\it vs}  the distance $l$ shown in Fig.\,4b yielded $\sigma_1 \approx 40$\,mJ/m$^2$. 
In combination with the average energy density obtained from the standard infinite wall configuration of Fig.\,1, we could see that $\sigma_2 \approx 130$\,mJ/m$^2$, which is indeed significantly larger than $\sigma_1$.

Why despite many prior computational studies of  domain walls the phenomenon of non-reciprocity escaped from the attention so far? 
One of the reasons is that the only non-reciprocal and highly symmetric wall in perovskite ferroelectrics is  this ($\bar{1}\bar{1}2$)-oriented 180$^{\circ} $ wall of rhombohedral phase. 
Moreover, embedding of ($\bar{1}\bar{1}2$) oriented wall requires a somewhat larger periodic supercell. 
%
The second reason is that the lowest-order coupling term introducing the non-reciprocity in Landau-Ginzburg theory is the third power of polarization gradient. 
This term has been systematically omitted in the earlier works.
%
At the same time, without a specific structural reason,  the non-reciprocal effects can be quite small. 
In case of BiFeO$_3$, the extraordinary enhancement mechanism consists in the coupling of the polarization to the oxygen octahedra tilt distortions by a third-order mixed polarization-tilt gradient term, combined with a Lifshitz-type bilinear coupling term \cite{tobe}.

In summary, we have explored the  180$^{\circ} $ domain walls of BiFeO$_3$ and find out that so far little investigated ($\bar{1}\bar{1}2$)-oriented  walls are the lowest energy among the R180$^{\circ}$ ones. We have given computational and symmetry arguments why these ($\bar{1}\bar{1}2$)-oriented walls are non-reciprocal and explained various profound consequences of this peculiar property.

This work was supported by the Czech Science Foundation (project no. 19-28594X) 
and by the European Union’s Horizon 2020 research and innovation programme 
under grant agreement no. 964931 (TSAR). 
MAPG acknowledge the European Union and the Czech Ministry of Education, Youth and Sports 
(Project: MSCA Fellowship CZ FZU I - CZ.02.01.01/00/22$\_$010/0002906).
Computational resources were provided by the e-INFRA CZ project (ID:90254), supported by the Ministry of Education, Youth and Sports of the Czech Republic. 
%


\end{document}